\documentclass[aip,onecolumn]{revtex4-1}

\usepackage{amsmath, amsthm, amssymb,amsfonts}
\usepackage[english]{babel}
\usepackage{color, graphicx}
\usepackage{epstopdf} 
\usepackage{comment}
\usepackage{caption}

\begin{document}

\title{North-South asymmetric Kelvin-Helmholtz instability and induced reconnection at the Earth's magnetospheric flanks}
\author{S.~Fadanelli}\email{sid.fadanelli@irap.omp.eu}\affiliation{Dipartimento di Fisica ``E. Fermi'', Universit\`a di Pisa, Largo B. Pontecorvo 3, 56127 Pisa, Italy}\affiliation{IRAP, Universit\'e de Toulouse, CNRS, CNES, UPS, (Toulouse), France}
\author{M.~Faganello}\email{matteo.faganello@univ-amu.fr}\affiliation{Aix-Marseille University, CNRS, PIIM UMR 7345, 13397, Marseille, France}
\author{F.~Califano}\email{francesco.califano@unipi.it}\affiliation{Dipartimento di Fisica ``E. Fermi'', Universit\`a di Pisa, Largo B. Pontecorvo 3, 56127 Pisa, Italy}
\author{S.~S.~Cerri}\email{scerri@astro.princeton.edu}\affiliation{Department of Astrophysical Sciences, Princeton University, 4 Ivy Ln, Princeton, NJ 08544, USA}
\author{F.~Pegoraro}\email{francesco.pegoraro@unipi.it}\affiliation{Dipartimento di Fisica ``E. Fermi'', Universit\`a di Pisa, Largo B. Pontecorvo 3, 56127 Pisa, Italy}
\author{B.~Lavraud}\email{benoit.lavraud@irap.omp.eu}\affiliation{IRAP, Universit\'e de Toulouse, CNRS, CNES, UPS, (Toulouse), France}

\begin{abstract}
  We present a three-dimensional study of the plasma dynamics at the flank magnetopause of the Earth's magnetosphere during mainly northward interplanetary magnetic field (IMF) periods. Two-fluid simulations show that the initial magnetic shear at the magnetopause and the field line bending caused by the dynamics itself (in a configuration taken as representative of the properties of the flank magnetopause) influence both the location where the Kelvin-Helmholtz (KH) instability  and  the induced magnetic reconnection take  place and their  nonlinear development. The KH vortices develop asymmetrically with respect to the Earth's equatorial plane where the local KH linear growth rate is maximal. Vortex driven reconnection processes take place at different latitudes, ranging from the equatorial plane to mid-latitude regions, but only in the hemisphere that turns out to be the less KH unstable. These results suggest that KH-induced reconnection is not limited to specific regions around the vortices (inside, below or above), but may be triggered over a broad and continuous range of locations in the vicinity of the vortices. 
\end{abstract}


\maketitle

\section{Introduction}\label{sec:intro}

The large-scale dynamics of the Earth's magnetosphere can be modelled, as a first approach, adopting a one-fluid magnetohydrodynamic (MHD) description. 
In fact  the magnetospheric plasma follows an ``ideal'' dynamics over most of its spatial domain, the magnetic field lines being frozen into the plasma motion and any cross-field diffusion being fairly negligible~\citep{SonnerupJGR1980,LabelleTreumannSSR1988,LeJGR1994}.
The magnetospheric region where field lines are ``anchored" to the Earth is separated from the heated solar wind plasma of the magnetosheath, where the interplanetary magnetic field (IMF) lines are connected to the open space, by a magnetic boundary known as the magnetopause.

Independently of its complex magnetic shape, in the absence of cross-field diffusion the frozen-in law prevents any kind of mixing between the magnetospheric and the solar wind plasmas. Therefore the plasma of solar wind origin could not in principle enter into the less dense magnetosphere. 

However, the frozen-in condition can be locally violated by non-ideal effects arising at small scales generated by the plasma dynamics itself, e.g. allowing for magnetic reconnection to occur. Since reconnection is capable of  modifying the global magnetic field topology, it strongly impacts both the dynamics of the whole system and the transport properties at the magnetopause. In particular during southward periods when the IMF direction is opposite to that of the magnetospheric magnetic field at low latitude, reconnection occurs at dayside magnetopause allowing for direct transport across the magnetopause and leading to the formation of a low-latitude boundary layer (LLBL) where solar wind and magnetospheric plasmas can mix~\citep{Dungey_1961}.

During northward periods the magnetic configuration at the low latitude magnetopause is unfavourable for magnetic reconnection to occur. 
Nevertheless the formation of a LLBL is observed also during  these periods~\citep{Mitchell_1987} up to the point that the entry of solar wind particles into the magnetosphere can be even more important than during southward periods~\citep{Terasawa_1997}. Different mechanisms have been invoked for explaining this transport which is routinely observed by satellites.

The non linear vortex dynamics resulting from the development of the Kelvin-Helmholtz (KH) instability is one of the few phenomena, together with lobe reconnection \citep{gosling_91,song_92,onsager_01} and kinetic Alfv\'en waves \citep{johnson_01a,johnson_01b,chaston_07}, able to explain the observed transport (see, e.g., \citet{FaganelloCalifano_2017} for further details on the respective role of the different phenomena). {The} KH instability is driven by the velocity shear between the stagnant magnetosphere and the flowing magnetosheath plasma of solar wind origin and grows along the magnetospheric flanks at low latitude, where the stabilizing magnetic shear is weaker for northward IMF. By contrast, under such northward IMF conditions, higher latitude regions are instead completely stabilized by the {stronger} magnetic shear.

{\it Per se}, the KH vortices developing during the non-linear phase  can strongly perturb the magnetopause but cannot mix the two different plasmas as their typical scale is so large that their early dynamics remains ``MHD-ideal''.
However, they become the driver of very fast secondary instabilities which give rise to a rich, small-scale non-linear dynamics that feeds on the energy source provided by the vortical motion: from secondary Kelvin-Helmholtz and Rayleigh-Taylor instabilities~\citep{Matsumoto_2004,Faganello_2008a,Tenerani_2011,Nakamura_2014}, to magnetic reconnection~\citep{KnollBrackbill_2002,Nakamura_2006,OttoFairfield_2000,Faganello_2008b,Faganello_2012b}, magnetorotational instability~\citep{Matsumoto_2007} or current-sheet kink instability~\citep{NakamuraPRL2004}.

If a magnetic shear exists across the low-latitude magnetopause, the KH velocity field will eventually pinch the magnetopause current sheet in between vortices and force the so-called ``Type I'' vortex induced reconnection to occur there~\citep{Liu_1988}. 
In this case reconnection must proceed on nearly the same ideal time-scale of the vortex dynamics in order to release the magnetic energy that piles-up at the compressed current sheet {carried} by the ideal motion~\citep{ChenJGR1997,KnollBrackbill_2002,Nakamura_2006,Nakamura_2013}.
Type I reconnection creates field lines that thread through the magnetopause, leading to a direct entry of solar wind particle into the magnetosphere. 

If the initial magnetic shear is set to zero and high-latitude KH stable region are included in the model, it has been shown that reconnection develops first at mid-latitude instead of around the equatorial plane where the vortices are generated by the primary KH instability. This process is driven by the braiding and the stretching of the field lines advected by the vortices at the equator but remaining anchored at higher latitudes in the Earth's magnetosphere~\citep{Faganello_2012b,Faganello_2014}.

Under such conditions, mid-latitude reconnection develops almost symmetrically with respect to the equatorial plane and creates double-reconnected flux tubes. These newly closed flux tubes, located on the Earthward side of the magnetopause, thus become populated with dense solar wind plasma. In this way, solar wind plasma enters the magnetosphere at a rate that is compatible with the observed one~\citep{Faganello_2012b}.

Recently, MMS spacecraft data have provided unambiguous in situ evidence of magnetic reconnection which were interpreted as Type I reconnection at the compressed current sheets forming in between primary successive KH vortices~\citep{Eriksson_2016}, confirming past observations with Cluster~\citep{nykyri_06,hasegawa_09}.
Remarkably, for the same MMS event evidences were also found for remote reconnection~\citep{Vernisse_2016}, i.e. occurring far away from the satellite location, as signalled by heated ions and electrons flowing parallel and anti-parallel along magnetic field lines just outside the magnetopause (e.g. \citet{gosling_91,fuselier_95,lavraud_06}.
These results suggest that Type I and mid-latitude reconnection coexist and cooperate in forming the LLBL for northward IMF, when a magnetic shear is present.

Here we present a numerical study  that takes  into account both a pre-existing shear between the magnetospheric field and the IMF, as well as the high-latitude stabilization of the KH instability, allowing for the simultaneous development of Type I and mid-latitude reconnection. 

In Sec.~\ref{sec:sim} we present the plasma model, the initial equilibrium and the parameters used in our simulations. 
\\
In Sec.~\ref{sec:KH} we show how the large-scale structures of the vortices are modified when both magnetic shear and high-latitude stabilization are present. 
\\
In Sec.~\ref{sec:rec} we present the analysis of the KH-induced reconnection processes. 
Finally, in Sec. \ref{sec:end} conclusions are drawn.


\section{Plasma model and simulation setup}\label{sec:sim}

We adopt a Hall-MHD plasma model (including finite resistivity). 
The model equations, in conservative form, are:
\begin{equation}
\partial_t n + \nabla \cdot (n {\bf u}) = 0 \label{eq:cont}
\end{equation} 
\begin{equation}
\partial_t (n {\bf u}) + \nabla \cdot \left(n {\bf u} {\bf u} + P_{tot} \bar{\bar{\bf I}} - {\bf B} {\bf B} \right) = {\bf 0} \label{eq:evol_u} 
\end{equation}
\begin{equation}
\partial_t {\bf B} = - \nabla \times {\bf E} \label{eq:evolv_B}
\end{equation}
\begin{equation}
{\bf E} = - {\bf u} \times {\bf B} + {\bf J}/n \times {\bf B} - \nabla P_e \, / n + \eta {\bf J} \label{eq:E} 
\end{equation}
where all quantities are normalized to ion (proton) quantities, the ion mass $m_i$, the inertial length $d_i$ and 
the Alfv\`en speed $v_A$. Here $n$ is the plasma number density, ${\bf u} \simeq {\bf u}_i$ the fluid velocity and $P_{tot}= P_i + P_e + B^2/2$. The ion and electron thermal pressures are evolved following an ideal adiabatic closure:
\begin{equation}\label{eq:closure}
\partial_t (nS_{i,e}) + \nabla \cdot \left(n S_{i,e} {\bf u}_{i,e}  \right) = 0 \ \ \ ; \ \ \ S_{i,e} = P_{i,e} n^{-5/3}
\end{equation}
Finally, we neglect the displacement current; then, the Faraday equation and the electron fluid velocity are given by
\begin{equation}\label{eq:J}
{\bf J} = \nabla \times {\bf B} \ \ \ ; \ \ \ {\bf u}_e = {\bf u} - {\bf J}/n
\end{equation}

With this model, during the initial large-scale dynamics leading to the formation of  fully rolled-up KH vortices, the magnetic field is frozen into the ion fluid motion and the dynamics is correctly  described by ideal MHD. 
During this phase the system spontaneously starts to distort and shrink the initial current sheet, eventually reaching a characteristic width comparable with the ion inertial length $d_i$. As a result, where the magnetic configuration is favourable, Hall-reconnection sets in on a fast time scale~\citep{Birn_2001,Faganello_2008c,Faganello_2012b}.
Admittedly, our model neglects the kinetic dynamics at scales smaller than $d_i$, as well as possible anisotropy effects~\citep[see, e.g.,][and references therein]{CerriPOP2013,DeCamillisPPCF2016}. Nevertheless, when implemented for reproducing a large portion of the boundary our model can realistically evaluate reconnection-induced plasma exchanges at the magnetopause~\citep[see][]{HenriPOP2013}.

The model equations are integrated numerically using a $4^{th}$ order Runge-Kutta scheme. Spatial derivatives are calculated using $6^{th}$ order explicit finite differences along the periodic $y$ and $z$ directions, while a $6^{th}$ order implicit compact scheme with spectral-like resolution~\citep{Lele_1992} is adopted along the inhomogeneous $x$-direction.
Very short wavelength fluctuations are dissipated using high order spectral filters acting only on the high-k part of the spectrum~\citep{Lele_1992}.

Special care is devoted to the boundaries along the inhomogeneous $x$-direction where we adopt transparent conditions for any MHD alfv\'enic or sonic perturbation generated inside the numerical domain. This method is based on projected characteristics of the ideal-MHD set of equations allowing one to control the in/out flux at the boundaries~\citep{Hedstrom_1979,Thompson_1990,Landi_2005,Faganello_2009}. In order to make this characteristic decomposition effective, buffer regions where non-ideal MHD terms are gradually smoothed out are implemented close to the $x$-boundaries.

\begin{figure}[bh]
\includegraphics[width=0.7\columnwidth]{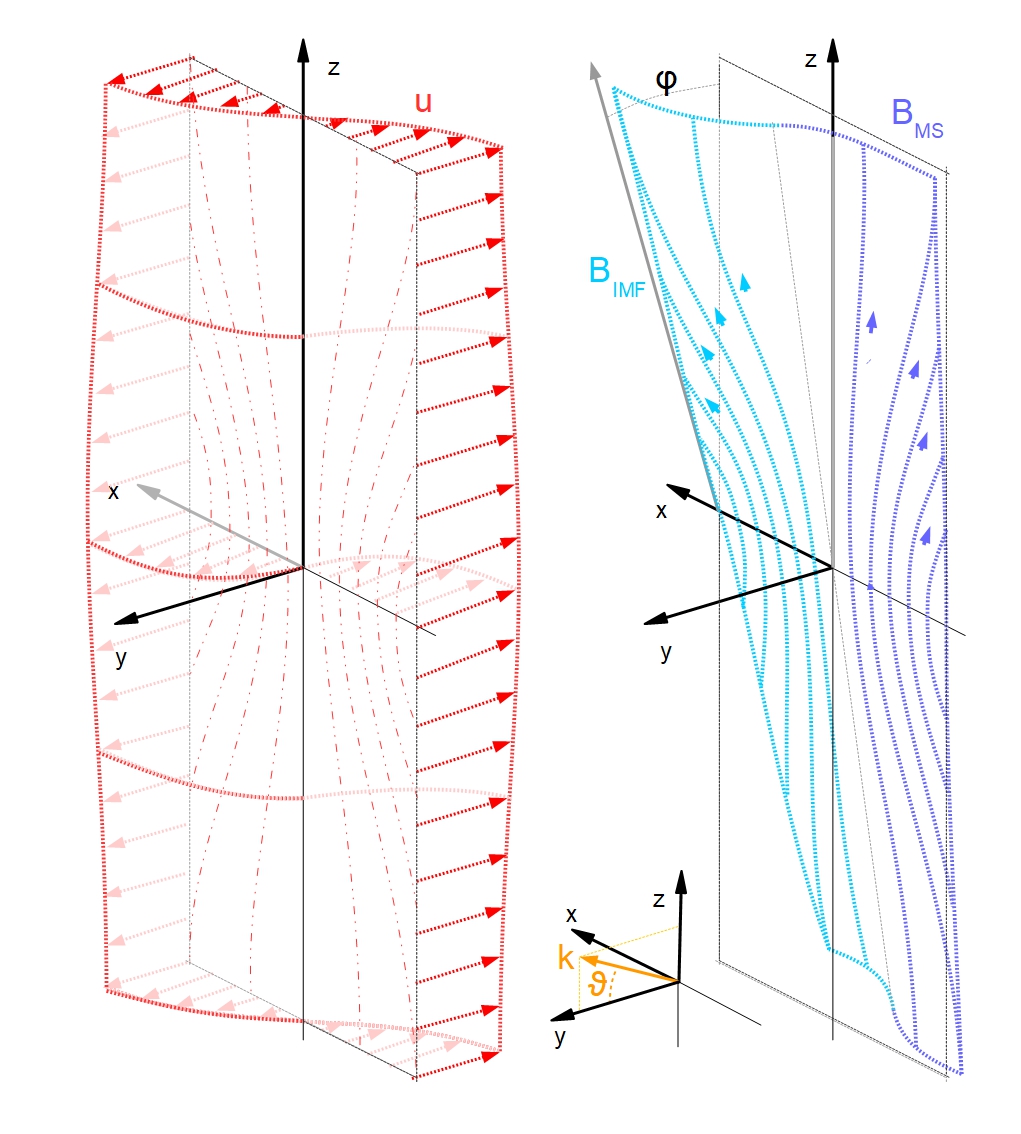}
\caption{Schematic representation of the magnetic and velocity fields in the equilibrium configuration. $\varphi$ represents the angle between the IMF and the northward direction ${\bf \hat{z}}$. $\vartheta$ defines the angle between a given wavevector ${\bf k}$ lying in the $(y,z)$-plane and the flow direction ${\bf \hat{y}}$.} \label{fig:equil_set}
\end{figure}

Simulations are initialized starting from a 2D ideal MHD equilibrium taken as uniform along the flow direction ($y$-coordinate). The $x$ and $z$ axis are set perpendicular to the unperturbed magnetopause and along the northward direction, respectively. In this configuration all equilibrium quantities are functions of $\psi$ only, where $\psi = \psi(x,z)$ is a magnetic flux function satisfying the Grad-Shafranov equations~\citep{AndreussiPOP2012,Faganello_2012b,Faganello_2012a}.
\begin{equation}\label{eq:GS}
\Delta \psi = \frac{d}{d\psi} \Pi \ \ \ ; \ \ \ \Pi = P_{0,i} + P_{0,e} + B_{0,y}^2/2
\end{equation}
Setting $\Pi = cst$ a simple solution is given by
\begin{equation}
\psi_0 (x,z) = \frac{1 + \delta}{2} x + \frac{1 - \delta}{2} \frac{L_z}{2 \pi} \, \sinh  \frac{2 \pi x}{L_z} \, \cos  \frac{2 \pi z }{L_z}  \label{eq:equil_psi}
\end{equation}
while the other equilibrium quantities are set as
\begin{equation}
n_0 = 1
\end{equation}
\begin{equation}
{\bf u}_0(x,z) = \frac{u_\star}{2} \tanh \frac{\psi_0(x,z)}{ \ell_\star} \, {\bf e}_y 
\end{equation}
\begin{equation}
{\bf B}_0(x,z) = \nabla \times \psi_0(x,z) {\bf e}_y + \frac{\tan (\varphi)}{2} \bigg[ 1 + \tanh \frac{\psi_0(x,z)}{ \ell_\star} \bigg] \, {\bf e}_y \label{eq:Beq}
\end{equation}
where $\varphi$ is the shear angle between the magnetospheric field and the IMF.
{The first term in  Eq.~(\ref{eq:Beq}) corresponds to the northward magnetospheric field ($x<0$) and the dominant northward component of the IMF ($x>0$). The second term adds a flow aligned component to the IMF, taking into account possible different configurations that are observed during periods of northward IMF. The equilibrium thermal pressure $P_{0,i} + P_{0,e}$ is the dominant term in $\Pi$ and varies from the magnetosphere to the magnetosheath in order to compensate for the increasing of $B_{0,y}^2/2$.}
A sketch of this equilibrium configuration is given in Fig.~\ref{fig:equil_set}.

In our simulations we set $L_z = 120\pi$, $\delta = 1/3$ and $\ell_\star = 3$ so that the equilibrium varies mainly along the $x$-direction and the velocity 
shear layer vorticity at $x=0$ is three times larger at $z=0$ then at $z=\pm L_z/2$.
As a consequence the KH instability, whose maximal growth rate is a fraction of the velocity shear layer vorticity~\citep{DrazinReid1981}, develops far faster in the equatorial region than at higher latitudes. This initial 2D configuration permits to mimic the preferential equatorial development of the KH instability at the flank magnetopause, under northward IMF. In the case of the Earth's magnetosphere, however, stabilization at higher latitudes is expected and observed owing to magnetosheath flow and magnetic field draping properties, so that magnetic and flow fields become more aligned and thus less prone to KH development~\citep{Chandrasekhar_1961}. The other two box dimensions are set as $L_x=90$ and $L_y=2\lambda_{FGM,z=0} = 30\pi$, where $\lambda_{FGM,z=0}$ is the expected wavelength of the fastest growing mode as given by a simplified 2D stability analysis at the equatorial plane. 

The sonic and alfv\'enic Mach number are defined as $M_s = u_\star/c_s$ and $M_A = u_\star / v_{A,z}$, where $c_s$ and $v_{A,z}$ are calculated at at the centre of the numerical box.  Their values, together with the other parameters, are listed in table \ref{tab:table1}. Finally, we take $P_i = P_e$ and $\eta = 0.001$.

\begin{table}[h]
\begin{tabular}{r|cc|c|l}
\hline
run &${M}_A$ &${M}_s$ &$\tan(\varphi)$ & description \\
\hline
``A" &$1.0$ &$\sqrt{3/5}$ &$0.3$  &weak magnetic shear\\
``B" &$2.0$ &$\sqrt{12/5}$ &$1.0$  &strong mag. shear \& high velocity\\
\hline
\end{tabular}
\caption{Summary of the relevant parameters characterizing the different simulations.}\label{tab:table1}
\end{table}

In order to follow the system evolution, to individuate the plasma structures forming during the dynamics and to follow the field line connectivity, we define a passive tracer $\varsigma$ advected by the fluid so as to mark the two different plasmas during the evolution. At the beginning of the simulation the passive tracer is set as
\begin{equation}\label{eq:pass}
\varsigma (x,z) = 0.6 + 0.4 \, \tanh \bigg[ \frac{\psi_0 (x,z)}{\ell_\star}\bigg]
\end{equation}
where $\varsigma<0.6$ {corresponds} to the magnetospheric plasma and $\varsigma>0.6$ to the solar wind one while $\varsigma \simeq 0.6$ determines {the position of }the magnetopause.
The passive tracer $\varsigma$ is constant along each magnetic field line and evolves as the field lines would do within ideal MHD. In this way $\varsigma$ {allows us}  to identify the reconnected lines linking the magnetospheric and solar wind plasma {as} those along which a variation of $\varsigma$ is measured.


\section{Large-scale dynamics of KH vortices}\label{sec:KH}

\subsection{Overview of the dynamics}

As expected, from the initial white noise perturbation {KH waves} emerge around the wavelength associated with the fastest-growing mode (FGM)  {as} predicted by linear theory. Given the length of the numerical box, two vortices appear at the end of the linear phase (not shown here).
As soon as they enter the non-linear phase the pairing process starts~\citep{winant_74,miura_97}.
As a result the vortices eventually merge generating a single larger vortex. 

\begin{figure}[t]
    \includegraphics[width=0.5\columnwidth]{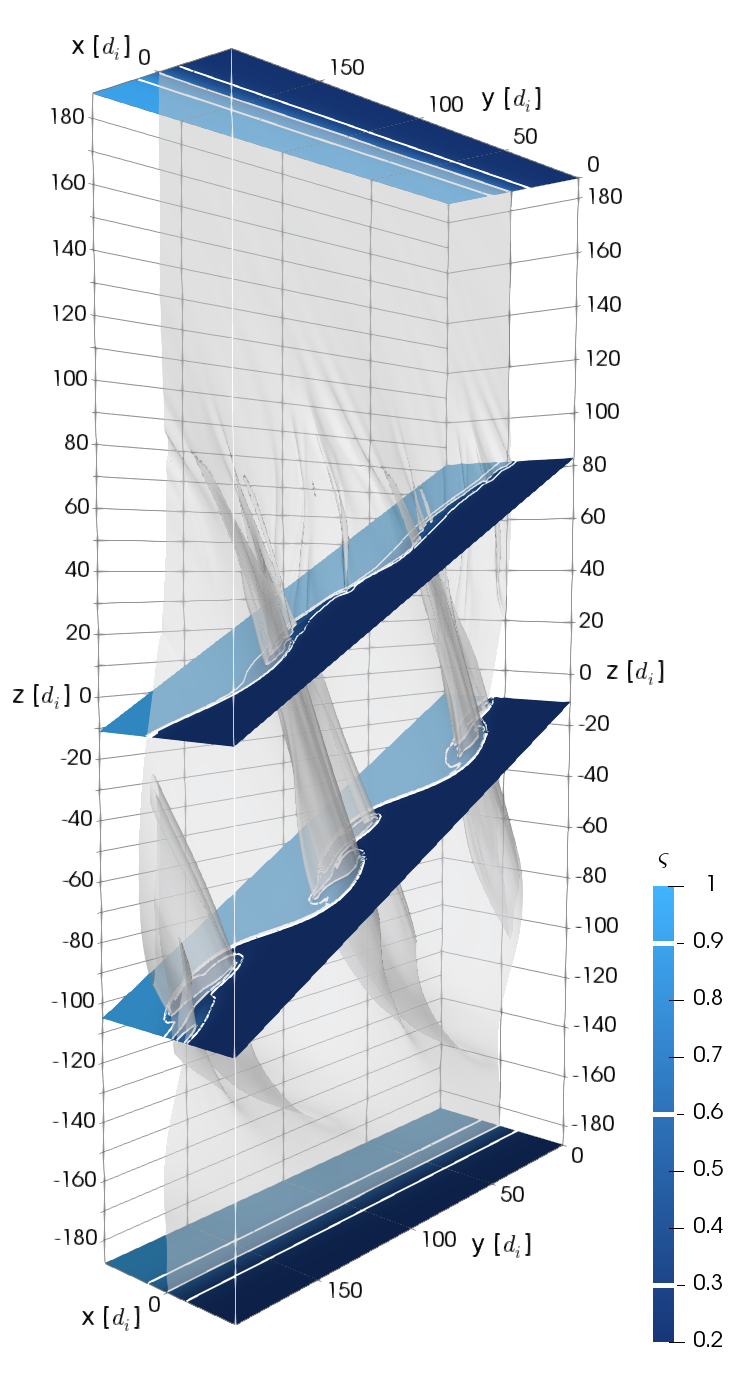}
    \caption{Visual rendering of the instability onset and development for $\tan \varphi = 1.0$ (``B" simulation) at $t=460$. The shaded isosurface ($\varsigma = 0.6$) corresponds to the magnetopause while dark/light blue colours correspond to the magnetospheric/solar wind plasmas. The white isocontours correspond to the passive tracer values $\varsigma = 0.3, 0.6, 0.9$. Note that for easing the visualization the box has been doubled along the $y$-direction.}  \label{fig:vis}
\end{figure}

In Fig.~\ref{fig:vis} we show the passive tracer iso-contours at $t = 460$  for run ``B". For sake of clarity, the box has been doubled along the $y$-direction so that two {pairs of coupling} vortices appear   instead of one. The semi-transparent quasi-vertical iso-surface corresponds to the magnetopause, $\varsigma = 0.6$, while the  {dark/light blue} colour correspond to the magnetospheric/solar wind plasma. 

Two pairing vortices have been produced around the equatorial region, the magnetopause being wrapped inside the vortex motions. The vortex structures, as shown by the folded magnetopause, extend both into the north and into the south hemispheres and are tilted with respect to the $z$-axis, corresponding to a KH wavevector not aligned with the initial flow. We recall that in the absence of an initial magnetic shear ($\tan \varphi = 0$), the KH vortex axis would be parallel to the $z$-axis. As expected for the chosen initial configuration, the vortices grow around the central region of the box while they are stable at high latitudes, as shown by the colour configuration in the unperturbed planes at $z=\pm L_z/2$.
However, the presence of an equilibrium magnetic shear breaks the reflection symmetry with respect to the equatorial plane of our initial configuration. Indeed while ${\bf u}_0$ and $B_{0,z}$ are symmetric, $B_{0,y} \rightarrow - B_{0,y}$ for $z \rightarrow -z$: the different properties under reflection follow from the fact that the velocity is a vector while the magnetic field is a pseudovector.

As a consequence, vortices develop differently in the northern and in the southern hemispheres, e.g. preferring the southern hemisphere for $\tan \varphi > 0$ as shown in Fig.~\ref{fig:vis} (the opposite is true for $\tan \varphi < 0$). The physical mechanism that favours one hemisphere with respect to the other is the  combined action of vortex growth and field line tying at high latitudes as will be discussed next in Sec.\ref{sec:KH_lat}.

\subsection{The tilting of unstable modes}\label{sec:KH_fou}

The vortex tilting observed in Fig.~\ref{fig:vis} is the consequence of the presence of a magnetic shear in the equilibrium configuration. This point can be understood as follows. The most unstable modes are the ones able to minimize the magnetic tension proportional to ${\bf k}\cdot{\bf B}_0$ (that counteracts the KH development) more than {to maximize} the driving term proportional to ${\bf k}\cdot {\bf u}_0$ (here ${\bf k} = 2\pi m/L_y {\bf e}_y + 2\pi n/L_z {\bf e}_z $ is the mode wavevector).
This effect has been proven to be at work when considering 1D equilibria varying only along $x$~\citep{Southwood_1968,Walker_1981}, but remains efficient also in our 2D equilibria with high-latitude stabilization. Indeed, the most unstable modes underlying the development of the vortex structures have a wavevector oblique with respect to the flow velocity and it is roughly perpendicular to the magnetic field direction (close to the velocity shear layer).

In order to calculate  the KH growth rate analytically, we consider the limit where the flow velocity and the magnetic field are uniform in two different regions separated by a sharp discontinuity at $x=0$. In our equilibrium configuration this would correspond to the limit $\ell_\star \rightarrow 0$ and $\delta \rightarrow 1$. By assuming incompressible perturbations, the KH growth rate is given by \citep{Chandrasekhar_1961}

\begin{equation}
  \gamma (k,\vartheta,M_A,\varphi) = k  \cos(\vartheta) \bigg[ \frac{{M}_A^2}{4} - \tan^2\vartheta - \tan\vartheta \,\tan\varphi - \frac{\tan^2\varphi}{2} \bigg]^{1/2} \label{eq:growth_rate}
\end{equation}
where $\vartheta$ is the angle between the wavevector and the flow direction ${\bf e}_y$. This system is unstable if and only if $\vartheta_-\leq\vartheta\leq\vartheta_+$, where $\vartheta_{\pm}$ is defined by 
\begin{equation}
\vartheta_{\pm} = \arctan \bigg[ - \frac{\tan\varphi}{2} \pm \frac{\sqrt{{M}_A^2 - \tan^2\varphi }}{2} \bigg] \label{eq:theta_interval} 
\end{equation}

The most unstable modes are thus found for 
\begin{equation}
2 \vartheta_{\max}  =  - \arctan\bigg( \frac{ 4 \tan \varphi}{4  - 2\tan^2 \varphi + {M}_A^2} \bigg) \label{eq:theta_max}
\end{equation}
that is different from zero provided that $\varphi \neq 0$. 
For small $M_A$ and small $\varphi$ we have  $\vartheta_{\max} \simeq - \varphi/2$, so that ${\bf k} \cdot {\bf B}_0 = 0$ around the center of the velocity shear layer. For large $\varphi$ the angle $\vartheta_{\max} < - \varphi/2$ because in our equilibrium configuration  the magnitude of ${\bf B}_0$ is bigger in the magnetosheath than in the magnetosphere by a factor $(1 + (\tan \varphi)^2)^{1/2}$. For large $M_A$ the stabilizing effect of the magnetic field is reduced so that the most unstable wavevector tends to be aligned with the flow ($|\vartheta_{\max}|$ decreases).
Note that for $z \rightarrow - z$, the magnetic shear angle $\varphi$ as well as $\vartheta$ change sign and that $\gamma(-\vartheta, -\varphi) = \gamma(\vartheta, \varphi)$.

\begin{figure}[th]
    \includegraphics[width=0.75\columnwidth]{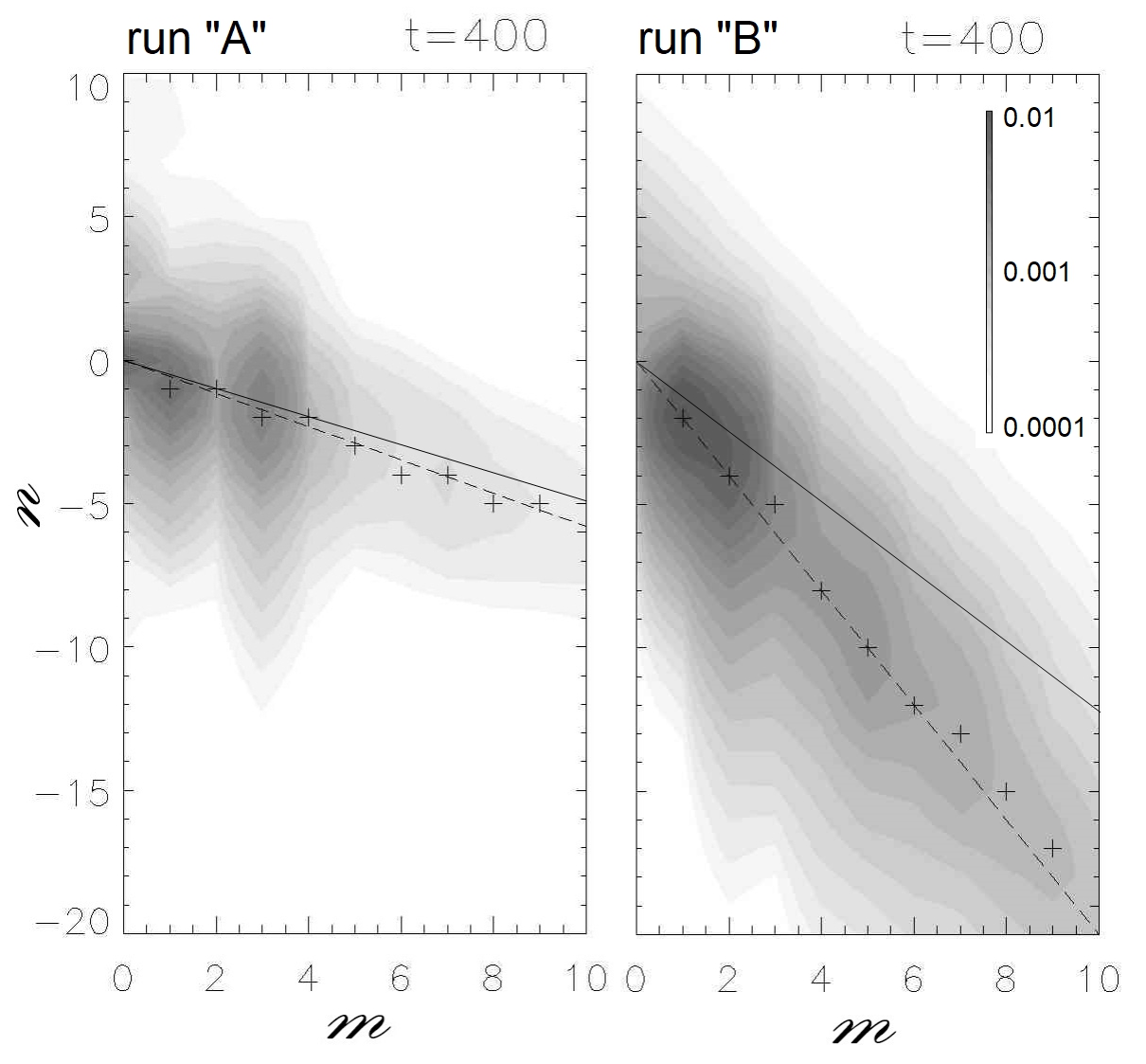}
    \caption{Shaded iso-contours of the $x$-averaged Fourier amplitude of $u_x$, normalised on the characteristic velocity $u_\star$, taken at $t=400$ for $\tan \vartheta = 0.3$ and $\tan \vartheta = 1.0$, left and right frames, respectively. The continuous lines represent the most unstable modes given by the (n,m) couples as predicted from the analytical step-like configuration. The dashed line by $\vartheta_{shear} = -\frac{1}{2} \tan \varphi$. For each discrete $m$-value, a cross indicates the location of maximal amplitude as obtained in simulations. Clearly crosses are almost aligned along the continuous line.}
\label{fig:KH_avx}
\end{figure}

Even if this model is oversimplified, it yet gives some insights about the tilt angle of oblique modes observed in the 3D compressible simulations starting from 1D equilibria with $\ell_\star \neq 0$ \citep{Nakamura_2014,Adamson_2016}. 
A moderate discrepancy between the predicted $\vartheta_{max}$ and that observed in the simulations is related to the fact that the simplified model underestimates the role of the magnetic field inside the shear layer ($|x| \lesssim \ell_\star$), where the mode amplitude is larger, while overestimates its importance in the two asymptotic region ($|x| \gg \ell_\star$). This is clearly shown by our simulations where the observed angle of the most unstable modes is slightly smaller than the predicted one.
The actual angle, for both run ``A" and ``B" is closer to $-0.5 \arctan \varphi$ than to $\vartheta_{max}$, even for $\varphi \sim 1$: the most unstable modes tend to develops perpendicular to the magnetic field at the center of the shear layer, minimizing the stabilizing role of the magnetic tension where the velocity shear term is stronger.
This is shown in Fig. \ref{fig:KH_avx} where we plot the magnitude of the $x$-averaged Fourier components of $u_x$ in the $(m,n)$-plane ($m$, $n$ are the mode numbers as defined before) for $\tan \varphi = 0.3$ and $\tan \varphi = 1.0$. For each $m$ number the largest amplitude correspond to $n<0$, i.e. to a tilted mode. The central region of the most unstable (tilted) modes observed in the simulations (gray strips) is aligned along the direction given by $\vartheta_{shear} = -\frac{1}{2} \tan \varphi$ (dashed line), so that ${\bf k}\cdot {\bf B}_0 \simeq 0$ at the center of the shear layer.
On the contrary  $\vartheta_{max}$ (continuous line) slightly underestimates the tilting.
For $\tan\varphi = 0.3$ we have $\vartheta_{max} \simeq 7^\circ$ and $\vartheta_{shear} \simeq 8^\circ$. For $\tan\varphi = 1.0$, $\vartheta_{max} \simeq 17^\circ$ and $\vartheta_{shear} \simeq 27^\circ$.


\subsection{Latitudinal shift of the vortices}\label{sec:KH_lat}

Due to the presence of the magnetic shear in the equilibrium configuration  the large-scale KH vortices  {extend asymmetrically with respect to} the equatorial plane. In particular, for $\tan \varphi > 0$, the latitude band  affected by the vortex structures shifts southward, below the equatorial plane. As we will discuss later, the opposite is true for $\tan \varphi < 0$.
\begin{figure}[ht]
    \includegraphics[width=0.75\columnwidth]{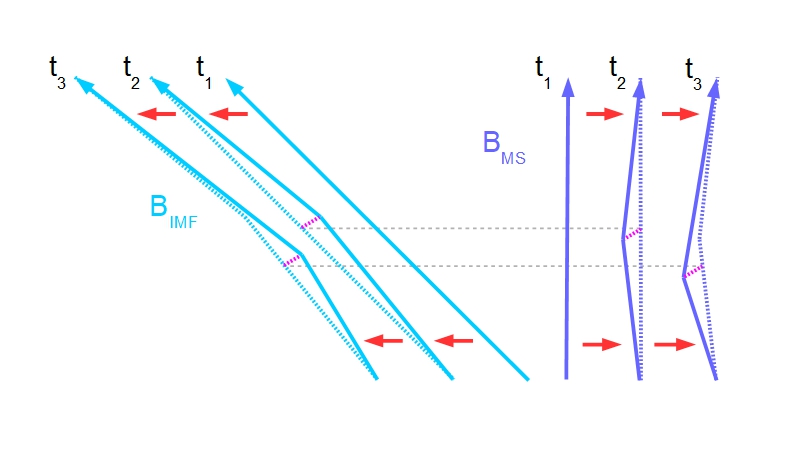}
\caption{A sketch of the differential magnetic field line advection mechanism for $\varphi > 0$. The figure shows a magnetospheric/IMF line, light blue and dark blue colours respectively, at three different times. The unperturbed field lines, denoted by $t_1$, first bend close to the equators, resulting into magnetic field lines at time $t_2$, due to the different advection at high/low latitudes, i.e. {field lines move unperturbed in opposite directions at high latitudes while they are slowed down in the equatorial plane.}
Indeed, as the field lines are frozen into the fluid motion, they are slowed down in the equatorial plane because they are embedded
in the vortex structures whose phase velocity is nearly zero. The magnetic shear is thus enhanced in the northern hemisphere and reduced in the southern one causing a southward drift of the instability. As a consequence the region where magnetic field field lines are slowed down gradually shifts southward, as shown for  $t = t_3$, as well as the region with smaller magnetic shear, favouring the KH development.}
\label{fig:diff_adv}
\end{figure}
Qualitatively this vortex shift can be explained by the differential advection of the magnetic field lines with respect to the latitude {position}. 

Differential advection has been discussed in the limit $\varphi  = 0$, i.e. zero magnetic shear, as an important driver for the magnetic field lines dynamics \citep{Faganello_2012b,Borgogno_2015}. Indeed, magnetic field lines  embedded  in the vortex structures {are slowed down in the equatorial region, with respect to their unperturbed motion, since the KH phase velocity is null in our frame. On the contrary they continue to move at the unperturbed magnetosphere/solar wind velocity at high latitudes.} As a consequence, magnetic field lines of  different origin are stretched and arched in the opposite directions, leading to the formation of intense  current sheets at mid latitudes where reconnection finally occurs.
When an initial magnetic shear is present, differential advection works {somewhat}  differently.  At the beginning the KH mode develops symmetrically with respect to $z=0$ but as soon as the vortices start to form, differential advection becomes more and more important and, contrary to the case without magnetic shear,  modifies the vortex structure  in a different way  above and below   the equatorial plane.

A sketch of this mechanism is given in Fig. \ref{fig:diff_adv} for $\varphi > 0$. We see that the initial magnetospheric and IMF lines, initially straight at $t=t_1$, are stretched by the differential advection, resulting in magnetic field lines that are increasingly bent at $t=t_2$ and $t=t_3$. As a consequence the magnetic shear is enhanced in the northern hemisphere while it is reduced in the southern one. Since the magnetic shear tends to inhibit the KH instability the location of the maximal growth rate gradually drifts southward. As a result, for $\varphi >0$, the KHI eventually develops faster in the southern hemisphere.

\begin{figure}[th]
\includegraphics[width=0.8\columnwidth]{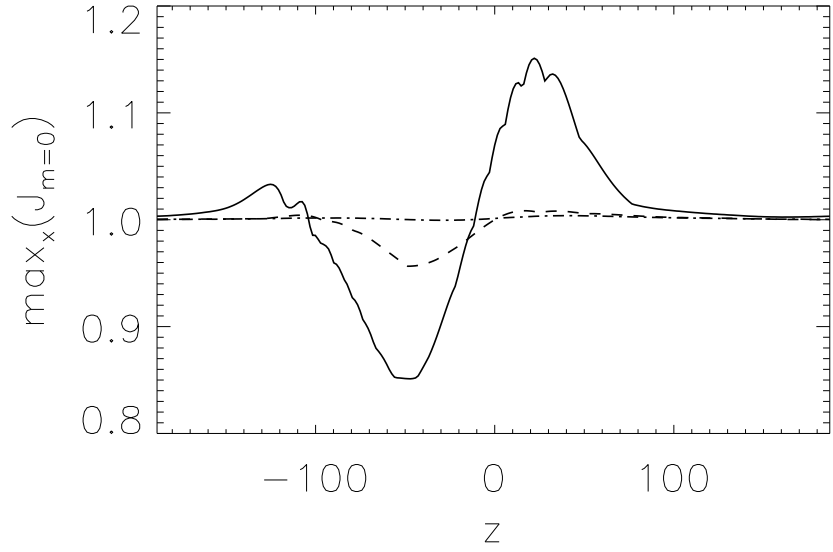} \\
\includegraphics[width=0.8\columnwidth]{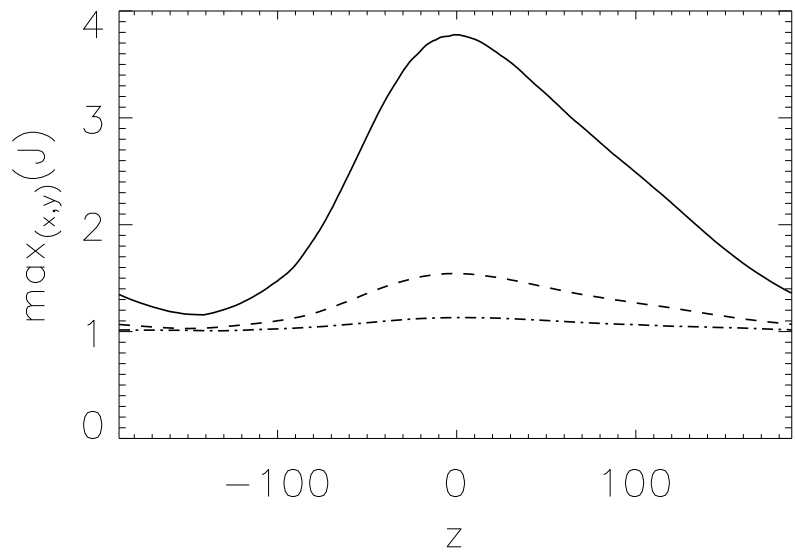} 
\caption{Top frame: $max_x (\tilde{J}_{m=0}(x,z))\, / \, max_x (J_0(x,z))$ as a function of $z$ at $t=250,300,350$. Bottom frame: $max_{(x,y)} (J(x,y,z))\, / \, max_x (J_0(x,z))$ at $t=250,300,350$.
}
\label{fig:shear}
\end{figure}

The different evolution of the magnetic shear in the two hemispheres can be quantified by looking at the peaks of electric current $J = |{\bf J}| = |\nabla \times {\bf B}|$. In Fig. \ref{fig:shear}, top frame, we plot $max_x (\tilde{J}_{m=0}(x,z))\, / \, max_x (J_0(x,z))$ as a function of $z$ for $t=250, 300, 350$, up to  the beginning of the nonlinear phase. Here $J_0$ is the magnitude of the equilibrium current while $J_{m=0}$ is the magnitude of the $m=0$ mode of the total current, thus including the nonlinear modification of the equilibrium.  It is clear that the $m=0$ magnetic shear is amplified in the northern hemisphere while it lowers in the southern hemisphere, explaining the southward shift of the KH unstable region at the beginning of the nonlinear phase.

In the present configuration, where the vorticity ${\bf \Omega}_0$ associated with the initial sheared flow is along $+{\bf \hat{z}}$, the equilibrium current ${\bf J}_0$ points out the hemisphere where the magnetic shear becomes larger. In general (for $\Omega_{0,z} \lessgtr 0$), the symmetry properties of the mechanism described in Fig.\ref{fig:diff_adv} suggest that differential advection enhances the magnetic shear in the northern hemisphere for ${\bf \Omega}_0 \cdot {\bf J}_0 > 0$, while the opposite is true for  ${\bf \Omega}_0 \cdot {\bf J}_0 < 0$. As a consequence KH vortices develop more vigorously in the southern/northern hemisphere for  ${\bf \Omega}_0 \cdot {\bf J}_0 \gtrless 0$.

As a reference, in Fig.\ref{fig:shear}, bottom frame, we plot the normalized value of the peaks of the total current $J = |{\bf J}|$ at $t=350$ as a function of $z$:  $max_{(x,y)} (J(x,y,z))\, / \, max_{x} (J_0(x,z))$. The maximal current increases at all latitudes due to the lateral compression of the original current sheet imposed by the KH velocity field. At the same time the current amplification is more important in the northern hemisphere as compared to the southern one because of differential advection.


\section{Magnetic field line dynamics}\label{sec:rec}

\subsection{Overview of the dynamics}

Most field lines maintain their connections during the whole dynamics even if  strongly bent and stretched by the KH vortical motion. In particular field lines on the left (right) of the magnetopause iso-surface $\varsigma = 0.6$ at $t=0$ {remain} on the same side. {On the other hand the connections of some field lines, such  as the yellow ones drawn in Fig. \ref{fig:rec}, are affected by magnetic reconnection occurring various places at the magnetopause. Now, these field lines connect two initially well separated magnetic domains, left and right of the magnetopause. This is shown in Fig.~\ref{fig:rec}, where such  magnetic field  lines cross the $\varsigma=0.6$ iso-surface at several  latitudes,  from  the magnetosheath (blue) to the magnetosphere (light blue), thereby connecting both sides of the magnetopause.

The dynamics investigated  here is more complex than that discussed in \citet{Faganello_2012b,Faganello_2012a,Faganello_2014,Borgogno_2015}. It includes at the same time a pre-existing magnetic shear between the magnetosheath and the  magnetospheric fields and high-latitude stabilization of the KH instability, so that reconnection can occur both as Type I or mid-latitude reconnection. The former process is driven by the pinching of the pre-existing current sheet caused by the compression in between KH vortices. Therefore it is expected to locally occur where the instability grows the most \citep{ChenJGR1997,KnollChaconPRL2002,Nakamura_2006}.
The latter is instead related to the field line differential advection and thus may be triggered at current sheets created (or modified) by this advection, far away from  the main location of the KH vortices  \citep{Faganello_2012b,Faganello_2012a,Faganello_2014,Borgogno_2015}.

We have shown in Fig.\ref{fig:shear}, bottom frame, that in the presence of a sheared magnetic field with $\varphi >0$ the combined action of differential advection and lateral compression increases the electric current at all latitudes but in particular in the northern hemisphere, i.e. in the hemisphere opposite to that  where the vortices are most intense. On this basis we may expect that Type I reconnection would preferentially occur around the equatorial region while mid-latitude reconnection would be favoured in the northern hemisphere.  In order to understand the development of such a complex  dynamics we need to determine a quantity that can act as a proxy for where reconnection occurs.


\subsection{Finding reconnection: a 3D diagnostic}\label{sec:MR_diag}

Determining reconnection sites in a full 3D, time-dependent geometry is far from straightforward. When a pre-existing current sheet is present, the current density $|{\bf J}|$ and  the magnetic shear already have quite ``large'' values, so they are not very useful when  seeking for reconnecting regions.   The passive tracer $\varsigma$  defined above (Eq.~(\ref{eq:pass})) is, instead, a convenient  proxy for defining ``reconnected'' field lines  since  only along these lines  a variation of $\varsigma$  can occur. However, such a tracer cannot  identify the precise location of ongoing reconnection}.

Hence, in order to find reconnection active {regions} we {define the following quantity: }
\begin{equation}
\varkappa = (\partial_t + {\bf u} \cdot \nabla)  ( S_e - S_i ) = ({\bf J} \cdot \nabla S_e ) /n\,,
\end{equation}
where Eqs.~(\ref{eq:cont}), (\ref{eq:closure}) and (\ref{eq:J}) have been used.
The idea behind this relies on the fact that {in our plasma model the entropy of each species is  passively advected by its respective fluid velocity. A difference between the electron and ion entropy advection thus indicate regions where ions and electrons decouple which,  in a Hall reconnection regime, include the reconnection regions. 
We thus expect that magnetic field lines passing through regions where {the value of $|\varkappa|$ peaks} are those undergoing reconnection. These may be either magnetic field that have not yet reconnected but that are advected by in-flows toward the center of the decoupling region, or magnetic field lines that have just reconnected and are moving away following the out-flows. Indeed, we observe in simulations that, as reconnection starts to act, the reconnected field lines highlighted using $|\varkappa|$ are those  with the the largest jump of $\varsigma$, with respect to that of several hundreds of randomly generated lines. Furthermore the value of the jump of these highlighted lines increases with time, i.e. as reconnection proceeds}.

\subsection{Latitudinal distribution of the reconnection processes}\label{sec:MR_lat}

\begin{figure}[t]
\includegraphics[width=0.47\columnwidth]{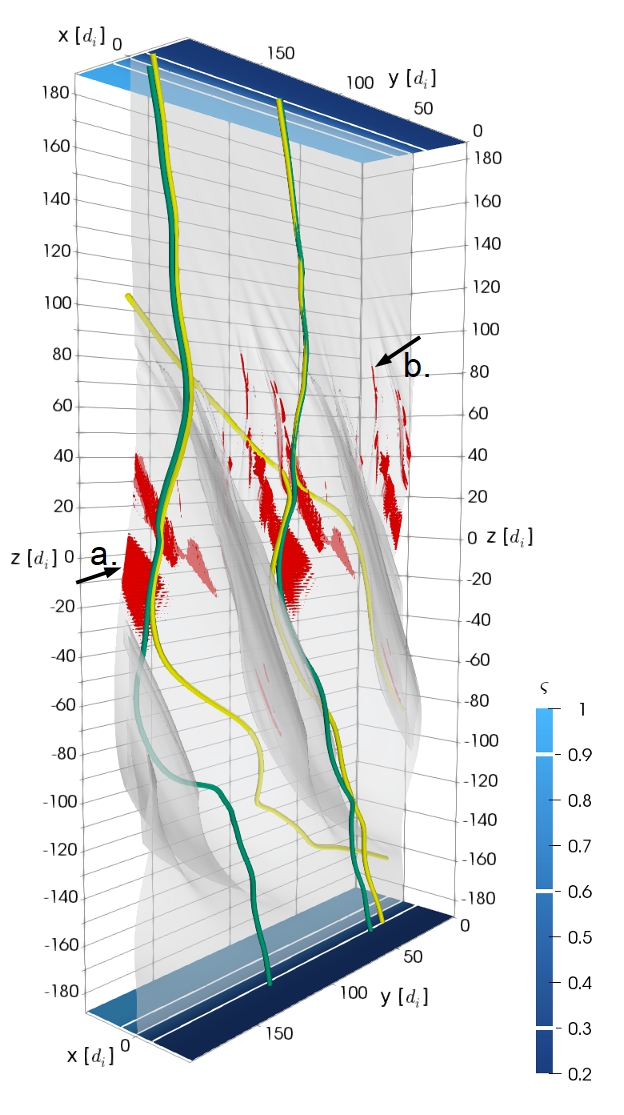} 
\caption{Shaded isosurface ($\varsigma = 0.6$) corresponding to the magnetopause and dark/light blue colours corresponding to the magnetospheric/magnetosheath plasmas for $\tan \varphi = 1.0$ at $t=460$. The white isocontours correspond to the passive tracer values $\varsigma = 0.3,  0.6, 0.9$. The regions where reconnection takes place are enclosed in the $|\varkappa|>0.03$ volumes, highlighted in red. The black arrows indicate planar (label ``a.'') and elongated (label ``b.'') reconnection sites. Some magnetic field lines representative of those crossing the active reconnecting sites have also been drawn - in yellow if reconnected, in green if not.
Note that for easing the visualization the box has been doubled along the $y$-direction. 
}
\label{fig:rec}
\end{figure}

In Fig.\ref{fig:rec} regions with large values of $|\varkappa|$ are shown as red surfaces  for simulation ``B" at $t=460$.  These active regions are all located  in the upper part of the latitude band  affected by the vortex structures, i.e. northwards with respect to the location where vortices are most intense. With respect to the vortex axis, active regions appear as large sheets in correspondence to  the hyperbolic points of the KH velocity field (``a.'' arrow) or as small scale filamentary structures aligned with the local magnetic field direction at the northernmost rippled boundary of the vortices (``b.'' arrow). 

When thought of in the frame of past works on the topic, these reconnection regions at the hyperbolic point  may be associated with either Type I or mid-latitude reconnection, since both {can} occur there even if at different latitudes.
However  the usual distinction between both types of reconnection {looses its meaning} when reconnection occurs over a large range of latitudes as observed here. We also note that the reconnecting regions appear as rather continuous patches from the latitude where vortices {are  most intense to the northernmost} end of the region affected by the KH instability.
The only distinction  concerns the mechanism by which reconnection is driven, i.e. how the current is enhanced. At the location where the vortices are most intense the magnetic shear grows mainly due to lateral compression, while in the northern regions it increases  mainly because of the magnetic field bending due to  differential advection. Therefore, even if a continuous set of reconnection sites is found ranging from one region to the other, we need to keep in mind  that the mechanism by which they have been triggered may not be the same.

Regarding the elongated reconnection regions, they are related to a small-scale rippling of the magnetopause, with a wavevector nearly perpendicular to the local magnetic field. This rippling appears at the northern edge of the region affected by the vortices and we conjecture that it is  related to a secondary instability that develops during  vortex pairing, namely the secondary KH instability~\citep{Cowee_2009,Matsumoto_2010,Tenerani_2011} or Type II magnetic reconnection, i.e. reconnection related to the folding of the flow-aligned component of the magnetic field that occurs during the pairing~\citep{Faganello_2008b,Faganello_2009}. In the former case it would be the velocity perturbations caused by the local ideal instability to cause reconnection~\citep{Tenerani_2011}. In the latter it would be reconnection itself to cause the plasma motion and thus the rippling. The detailed analysis of secondary instabilities and induced reconnection is beyond the scope of this paper and will be tackled in a future work. Nevertheless contrary to what observed in 3D simulations neglecting high-latitude stabilization~\citep{Nakamura_2013,Nakamura_2014}, in our simulations secondary instabilities occurs far away from the region  where the primary KH vortices are more intense.

\subsection{Double reconnection processes}\label{sec:MR_double}

We define double-reconnected field lines as those lines that  undergo  reconnection twice  at different latitudes. About half of these lines connect the magnetosheath in the equatorial region to the magnetospheric plasma at high latitudes (and viceversa). They are particularly important, as compared to once-reconnected lines that simply ``open'' the magnetopause (allowing for the development of an open LLBL) because they can effectively trap solar wind plasma onto closed field lines of the magnetosphere. Indeed, the flux tubes associated to these lines can be considered as new magnetospheric flux tubes with their low-latitude portion populated by solar wind particles.
Also, the creation of double-reconnected flux tubes can explain the {increase} of the specific entropy of the cold ion population measured just inside the magnetopause \citep{JohnsonWingJGRA2009}. {Indeed, a statistical survey of the low-latitude magnetosphere during northward periods  shows that the cold dense population of the magnetosheath leaks through the magnetopause  increasing its specific entropy by a factor $5 \div 20$.}

In the absence of a pre-existing shear between the IMF and the magnetospheric field double reconnection involves two locations  along the same field line in the two opposite hemispheres, acting in the northern as well as in the southern hemisphere in a nearly symmetric way. Adding a magnetic shear to the system not only breaks the symmetry but  changes where and how reconnection develops. Recent MMS data show that reconnection occurs in the region  where the vortices are observed and also far away from the vortex location, possibly at mid-latitude \citep{Vernisse_2016}.  
Our simulations confirm this scenario, showing that for a positive magnetic shear angle ($\varphi>0$) reconnection occurs at the same time  where the vortices are most intense and in regions that are northern that this latitude
 (the opposite is true for a negative magnetic shear angle,  $\varphi < 0$). In addition our simulations show that double-reconnected lines are generated during the late non-linear phase of the vortex dynamics for both $\tan\varphi=0.3$  and $\tan\varphi=1.0$. 
 

\section{Conclusions}\label{sec:end}

We have investigated the  development of the Kelvin-Helmholtz instability and the induced reconnection processes in a geometry that models the configuration of the flanks of the Earth's magnetosphere during periods of northward IMF by means of high-resolution Two-fluid simulations.
Our initial configuration takes into account both the effect of high-latitude stabilization and  of a pre-existing magnetic shear between the magnetospheric and the magnetosheath fields.  
The most remarkable features of the plasma dynamics observed in 
this configuration are  the latitude location where the KH instability grows more vigorously, the place where induced magnetic reconnection occurs and  the mechanism underlying induced reconnection.

Concerning the first point, as soon as the IMF has a component along the flow (described by a shear angle $\varphi$ in our simulations) the reflection symmetry about the equatorial plane is broken even if the density, temperature and velocity field are symmetric (the northward component of the magnetic field is symmetric too).
In particular, we have shown that KH vortices develop asymmetrically with respect to the equatorial plane depending  on the sign of the pseudoscalar ${\bf \Omega}_0\cdot{\bf J}_0$, where ${\bf J}_0$ is the equilibrium  current associated to the rotation of the equilibrium magnetic field across the magnetopause and ${\bf \Omega}_0$ is the equilibrium vorticity associated to the velocity shear.
When ${\bf \Omega}_0\cdot{\bf J}_0 > 0$, KH vortices fully develop mainly in the southern hemisphere, whereas the contrary is true for ${\bf \Omega}_0\cdot{\bf J}_0 < 0$.

From a physical point of view the shift of the vortices towards  one hemisphere can be explained by looking at the dynamics of magnetic field lines at the beginning of the nonlinear phase. In fact, even if the linear KH growth rate is symmetric, the dynamics of field lines is not. Indeed, the magnetic field lines are frozen in the plasma fluid motion and are advected differently at high latitudes, where the magnetospheric/solar wind velocity stays unperturbed, and at low latitude where the instability develops. This differential advection causes the averaged magnetic field shear to increase in one hemisphere and to be reduced in the other one. Since the magnetic shear tends to inhibit the KH growth, vortices develop asymmetrically.

Since both ${\bf \Omega}_0$ and ${\bf J}_0$ change sign when passing from the magnetospheric dusk flank to the dawn flank, the hemisphere where the vortices are more intense is the same at both flanks, e.g. the southern one if the flow-aligned component of the IMF is positive. This fact can be directly inferred from the symmetry properties of the system: the dawn flank configuration can be obtained from the dusk one by reflecting the system with respect to the magnetopause and applying charge-conjugation. Since MHD equations are invariant under ``reflection + charge-conjugation'', the large-scale KH dynamics is specular.

In the past, \citet{farrugia_98} and \citet{gratton_03}  considered the impact of the clock angle of the IMF on the KH instability, i.e. the impact of a westward component of the magnetic field, perpendicular to both the northward and the flow directions. Neglecting the flow-aligned component of the IMF, they showed that for a positive clock angle the location of the maximum linear growth rate of the KH instability is located in the northern hemisphere at the dusk side. The opposite is true at the dawn flank so that the most unstable hemispheres are different at the dawn/dusk sides. This behaviour has been obtained by looking at the configuration of the magnetic field close to the global magnetopause, taking into account the dipolar configuration of the magnetospheric field and the draping of the solar wind magnetic field around the magnetopause as described by a global MHD code.
From a symmetry point of view this fact is not surprising since as soon as a westward component of the magnetic field is considered the global system is no more invariant under ``reflection + charge-conjugation'' so that the dawn and dusk large-scale dynamics are not specular (here the reflection of the global system is about the plane defined by solar wind direction and the northward direction, passing through the Earth).

In our configuration the clock angle is not included so that the nonlinear KH activity at the flanks is specular. On the contrary when considering the clock angle but neglecting the shear angle \citep{farrugia_98,gratton_03} the linear dynamics is anti-specular. Taking into account both
the shear angle and the clock angle effect would help in clarifying  satellite data analysis, in particular when  KH activities measured at both flanks (at different latitudes) are compared \citep{hasegawa_06,nishino_11,taylor_12}.

The magnetic shear angle has a similar but opposite impact on the location where reconnection occurs, with respect to the location where the KH vortices eventually settle, since reconnection develops faster in regions where the magnetic shear is larger. The local magnetic shear is enhanced in two different ways during the nonlinear dynamics. The first one is the pinching of the pre-existing current sheet that occurs at the hyperbolic point (in between successive vortices) of the KH velocity field, as in Type~I reconnection. 
The second one is the modification of the pre-existing current sheet far away from the location where vortices are intense due to differential advection and field line bending. The local magnetic shear becomes larger in the hemisphere opposite to the one where KH vortices are more developed, i.e. in the northern (southern) hemisphere for a positive (negative) ${\bf \Omega}_0 \cdot {\bf J}_0$. Both mechanisms are at work in our simulations, leading to the development of reconnection  in a wide latitude range: from the location where vortices are most intense to the upper (lower) edge of the region affected by KH perturbations.

Recent MMS observations suggest that that Type I reconnection proceeds nearby vortices~\citep{Eriksson_2016} and that, at the same time, remote reconnection occurs probably at mid latitudes~\citep{Vernisse_2016}. The  simulations discussed here  reproduce this dynamics and and further suggest that remote reconnection should be favoured in a given hemisphere depending on the initial magnetic shear (i.e., the prevailing IMF orientation). In particular when the flow-aligned component of the IMF is negative, as during the MMS observations, the favoured hemisphere is the southern one. This is compatible with the fact that the number of remote reconnection events in MMS data is bigger in the southern than in the northern hemisphere~\citep{Vernisse_2016}.

Our simulations also show that reconnection, going on at different latitudes, is able to produce double-reconnected magnetic field lines connected to the Earth and thus to trap dense magnetosheath plasma inside the magnetopause even when a significant magnetic shear is present in the initial configuration. 
These results indicate that this double reconnection process associated with KH vortices is a viable mechanism to explain the formation of the flank LLBL even in the presence of significant magnetic shear. Future work shall focus on determining the efficiency of this mechanism as a function of the initial magnetic shear. 

\acknowledgments

The simulations presented here have been performed at CINECA (Bologna, Italy) under the ISCRA allocation initiative.



%

\end{document}